\documentclass[11pt]{article}
\usepackage[textwidth=15.2cm,textheight=22cm]{geometry}
\usepackage{amsmath,amssymb}
\usepackage{latexsym}
\usepackage{multicol}
\usepackage{graphicx}
\usepackage{caption}
\usepackage{scalerel}
\usepackage{bm}
\numberwithin{equation}{section}
\pdfoutput=1
\usepackage[T1]{fontenc}
\usepackage[latin9]{inputenc}
\usepackage[active]{srcltx}
\usepackage{esint}
\usepackage{cancel}
\usepackage{color}
\usepackage{ulem}
\allowdisplaybreaks[1]
\def\bh{\bar h}
\def\bp{\bar \partial}
\def\p{\partial}
\def\parp{\partial_+}

\def\beas{\begin{eqnarray*}}
\def\eeas{\end{eqnarray*}}

\newcommand{\del}{\partial}
\newcommand{\be}{\begin{equation}}
\newcommand{\ee}{\end{equation}}
\newcommand{\bea}{\begin{eqnarray}}
\newcommand{\eea}{\end{eqnarray}}

\newcommand{\half}{\frac{1}{2}}
\newcommand{\nn}{\nonumber}

\newcommand{\ndt}{\noindent}

\newcommand{\parm}{\partial_-}

\newcommand\fr[1]{\frac{1}{#1}}

\definecolor{ggreen}{rgb}{0.0, 0.5, 0.20}

\begin{document}

	\begin{titlepage}
		\begin{flushright}    
			{\small $\,$}
		\end{flushright}
		\vskip 1cm
		\centerline{\Large{\bf{The structure of interaction vertices}}}
		\vskip .5cm
		\centerline{\Large{\bf{in pure gravity in the light-cone gauge}}}
		\vskip 1.5cm
		\centerline{Sudarshan Ananth$^\dagger$, Nipun Bhave$^\dagger$ and S.I Aadharsh Raj$^*$}
		\vskip .7cm
		\centerline{$\dagger$\,\it {Indian Institute of Science Education and Research}}
		\centerline{\it {Pune 411008, India}}
		\vskip 0.3cm
		\centerline{$^*\,$\it {UM-DAE Centre for Excellence in Basic Sciences}}
		\centerline{\it {Mumbai 400098, India}}
		\vskip 1.5cm
		\centerline{\bf {Abstract}}
		\vskip .5cm
		
		\ndt The first truly non-MHV interaction vertices in the light-cone formulation of pure gravity appear at order 6.  From a closed form expression, for gravitation in the light-cone gauge, we extract and present all 6-point interaction vertices. We invoke symmetry arguments to explain the structure of these vertices. Symmetry considerations also allow us to place constraints on the structure of all even- and odd-point vertices in the theory. The origin of MHV vertices within this formalism is also discussed. 
		
		\vfill
	\end{titlepage}

	\section{Introduction}
\ndt Neither locality nor Poincar{\' e} invariance is manifest in theories formulated in the light-cone gauge. While this may appear to be a limitation, it serves as a significant advantage when dealing with scattering amplitudes. This is because both these properties, when manifest, obscure the inherent simplicity underlying amplitude structures (as does the standard covariant formalism). For example, the Parke-Taylor formula for n-gluon scattering~\cite{Park} emerges naturally from the light-cone formulation of Yang-Mills theory.
\vskip 0.3cm

\ndt There is a price to pay for rendering the simplicity manifest - light-cone formulations are significantly more involved than their covariant counterparts. Gravity in the light-cone gauge was studied in~\cite{Scherk:1974zm, BCL}. The primary result being a closed form expression for the gravitational Lagrangian, written entirely in terms of the two physical degrees of freedom in the theory. Subsequently, this framework was further refined and perturbative expansions of the closed form Lagrangian achieved to orders $\kappa$, $\kappa^2$ and $\kappa^3$~\cite{Ananth:2006fh,Ananth:2008ik}. The close links between the light-cone gauge and spinor helicity variables~\cite{Ananth:2012un} ensures that the standard tree-level amplitudes emerge naturally from these perturbative results (in momentum space).

\vskip 0.3cm

\ndt A limitation of existing perturbative studies is their focus on MHV structures~\cite{CSW}. In the light-cone gauge, all quartic vertices for massless bosonic fields, are inherently MHV. Five-point interaction vertices do involve structures having three negative helicity fields and two positive helicity fields, but these vertices are essentially the conjugates of 5-point MHV vertices. Thus the first truly non-MHV vertices appear at the six-point level. This is the focus of the first part of this paper. After reviewing the closed form expression for the light-cone gravity action, we explicitly derive all the 6-point interaction vertices in the theory. The second part of the paper focuses on the symmetries in the theory and the constraints they impose on `allowed' interaction vertices. The six-point vertices, for example, are all non-MHV in structure and we explain why this is the case from simple symmetry considerations. These symmetry arguments extend to all orders and determine the nature of even- and odd-point interaction vertices in the light-cone formulation of pure gravity. We also comment on the origin of MHV vertices in the theory.

	\section{$3-$, $4-$ and $5-$point interaction vertices in light-cone gravity: a review}
	
\ndt With the metric $(-,+,+,+)$, we define the light-cone coordinates
	\bea \label{coord}
	x^\pm=\fr{\sqrt 2}(x^0\pm x^3)\ ,&& x=\frac{1}{\sqrt 2}\,(\,{x_1}\,+\,i\,{x_2}\,)\ , \qquad \bar x= x^*\ .
	\eea
	$\partial_\pm, \bar\partial$ and $\partial$ are the corresponding derivatives. $x^+$ is chosen as light-cone time. So $-i\partial^-$ is the light-cone Hamiltonian. 
\vskip 0.1cm
	\ndt The Einstein-Hilbert action on a Minkowski background is
	\bea
	S_{EH}=\int\,{d^4}x\;\mathcal{L}\,=\,\frac{1}{2\,\kappa^2}\,\int\,{d^4}x\;{\sqrt {-g}}\,\,{\mathcal R}\ ,
	\eea
	where $g=det\,(\,{g_{\mu\nu}}\,)$ is the determinant of the metric. $\mathcal R$ is the curvature scalar and $\kappa^2=8\pi\, G$ is the gravitational coupling constant. The corresponding field equations are
	\bea
	\label{feq}
	\mathcal R_{\mu\nu} \,-\,\half\, g_{\mu\nu}\mathcal R\,=\,0\ .
	\eea
At this stage, we impose {\it {three}} gauge choices~\cite{Scherk:1974zm, BCL}
	\bea \label{lcg}
	g_{--}\,=\,g_{-i}\,=\,0\quad ,\; i=1,2\ ,
	\eea
	\ndt and parameterize the other components as 
	\bea
	\label{gc}
	\begin{split}
		g_{+-}\,&=\,-\,e^\phi\ , \\
		g_{i\,j}\,&=\,e^\psi\,\gamma_{ij}\ .
	\end{split}
	\eea
	The matrix $\gamma^{ij}$ is real, symmetric and unimodular. $\phi$ and $\psi$ are real parameters. Field equations in which $\del_+$ (time derivative) does not appear serve as constraint relations, as opposed to `true' equations of motion. The constraint relation $\mu\!=\!\nu\!=\!-\;$ from (\ref {feq}) reads
	\bea \label{CE1}
	2\,\del_-\phi\,\del_-\psi\,-\,2\,\del^2_-\psi\,-\,(\del_-\psi)^2\,+\, \half \del_-\gamma^{ij}\,\del_-\gamma_{ij}\,=\,0\ .
	\eea
	This can be solved by making the specific choice  
	\bea 
           \label{fourth}
	\phi\,=\,\frac{\psi}{2}\ ,
	\eea 
which is the {\it {fourth}} (and final) gauge choice. With this choice 
	\bea 
	\label{psio}
	\psi\,=\,\frac{1}{4}\,\frac{1}{\del^2_-}\,(\del_-\gamma^{ij}\,\del_-\gamma_{ij})\ .
	\eea 
	Unphysical components of the metric are systematically eliminated by constraint relations. This leaves us with a `physical' closed-form expression~\cite{Scherk:1974zm,BCL} entirely in terms of the $\gamma^{ij}$. In the expressions below, we continue to use $\phi$ \& $\psi$ (even though (\ref {fourth}), (\ref {psio}) relate them to the $\gamma^{ij}$) because they make terms in the calculation easy to keep track of. 
	\bea
	\label{aaction}
	S\,&=&\frac{1}{2\kappa^2}\int d^{4}x \; e^{\psi}\left(2\,\del_{+}\del_{-}\phi\, +\, \del_+\del_-\psi - \half\,\del_{+}\gamma^{ij}\del_{-}\gamma_{ij}\right) \nonumber \\
	&&-e^{\phi}\gamma^{ij}\left(\del_{i}\del_{j}\phi + \half \del_{i}\phi\del_{j}\phi - \del_{i}\phi\del_{j}\psi - \frac{1}{4}\del_{i}\gamma^{kl}\del_{j}\gamma_{kl} + \half \del_{i}\gamma^{kl}\del_{k}\gamma_{jl}\right) \nn \\
	&&- \half e^{\phi - 2\psi}\gamma^{ij}\frac{1}{\del_{-}}R_{i}\frac{1}{\del_{-}}R_{j}\ ,
	\eea
	\ndt where 
	\bea 
	R_{i}\,\equiv\, e^{\psi}\left(\half \del_-\gamma^{jk}\del_{i}\gamma_{jk}-\del_-\del_i\phi - \del_-\del_i\psi + \del_i\phi\del_-\psi\right)+\del_k(e^\psi\,\gamma^{jk}\del_-\gamma_{ij})\ . \nn
	\eea

\vskip 0.5cm
\subsection{Perturbative expansion}
\vskip 0.3cm

\ndt In~\cite{BCL}, a perturbative expansion of the Lagrangian in (\ref {aaction}), in powers of $\kappa$, was obtained using
\bea
\label{gamma}
\gamma_{ij}=\left(\mathrm{e}^{ \sqrt{2}\;\kappa \, h\, }\right)_{ij}\ ,\qquad \gamma_{ij}\,\gamma^{jk}=\delta_i^k \;\;,\qquad h_{ij}=\begin{pmatrix} h_{11} & h_{12}\\h_{12} &-h_{11}\end{pmatrix}\ .
\eea
These relate to the physical helicity states of the graviton as
\bea
\label{grav}
h_{ij}=\fr{\sqrt 2}\begin{pmatrix} h+\bh & -i(h-\bh)\\-i(h-\bh) &-h-\bh\end{pmatrix}\ ,
\eea
with $h$ and $\bh$ representing gravitons of helicity $+2$ and $-2$ respectively. 
\vskip 0.2cm
\ndt Instead of working with the $h_{ij}$ and then rewriting quantities in terms of helicity states, we focus directly on the helicity states by introducing transverse variables $a$ and $b$ (so $x^a\,, x^b$ represent $x\,,\bar x$)~\cite{Ananth:2008ik}. The relevant metric being
\bea
\label{transf}
\eta_{ab}=\frac{\p x^i}{\p x^a}\,\frac{\p x^j}{\p x^b}\,\eta_{ij}\ .
\eea
So
\bea
\eta_{xx}=\eta_{\bar x\bar x}=0\ ,\qquad \eta_{x{\bar x}}=\eta_{{\bar x}x}=1\ .
\eea
The inverse metric ($\eta^{ab}\eta_{bc}=\delta^a_{\;\;c}$) has components
\bea
\eta^{xx}=\eta^{\bar x\bar x}=0\ ,\qquad \eta^{x{\bar x}}=\eta^{{\bar x}x}=1\ .
\eea
In this new basis, the expressions for $\gamma$ from (\ref {gamma}) read
\bea
\label{matrixone}
\gamma_{ab}\,\,=\,\, \sum_{n = 0}^{\infty} \begin{pmatrix} 2\kappa^{2n+1}\,\dfrac{(4h\bar{h})^n}{(2n+1)!}\,\bh & \kappa^{2n}\dfrac{(4h\bar{h})^n}{(2n)!}\, \\\\ \kappa^{2n}\dfrac{(4h\bar{h})^n}{(2n)!}\, & 2\kappa^{2n+1}\,\dfrac{(4h\bar{h})^n}{(2n+1)!}\,h \end{pmatrix}\ ,
\eea
\bea
\label{matrixtwo}
\gamma^{ab}\,\,=\,\,\sum_{n = 0}^{\infty}\begin{pmatrix}  -2\kappa^{2n+1}\,\dfrac{(4h\bar{h})^n}{(2n+1)!}\,h & \kappa^{2n}\dfrac{(4h\bar{h})^n}{(2n)!}\, \\\\ \kappa^{2n}\dfrac{(4h\bar{h})^n}{(2n)!}\,& -2\kappa^{2n+1}\,\dfrac{(4h\bar{h})^n}{(2n+1)!}\,\bh  \end{pmatrix}\ .
\eea
For the purposes of this paper, we only need terms up to $n=2$ in the series (\ref{matrixone}) and (\ref{matrixtwo}). From these, we can also calculate $\psi$ at orders $2$ and $4$
\bea
\psi\!\!\!\! && = \qquad 2\kappa^2 \psi_2 \qquad \;\;\;\;\;\;\;\;\;\,\;+\;\;\;\;\;\; \qquad 4\kappa^4 \psi_4  \\
&&=2\kappa^2\left\{ \!-\fr{\parm^2}(\parm \bh\parm h)\right\} + 4\kappa^4\left\{- \fr{3\parm^2}(\parm h\parm[\bh\bh h])\!-\! \fr{3\parm^2}(\parm[\bh hh]\parm\bh)\!+\!\fr{2\parm^2}(\parm[\bh h]\parm[\bh h])\ \right\}\nn\ .
\eea
\vskip 0.3cm
\ndt We now expand the closed form action in (\ref {aaction}) to order $\kappa^2$. This reads~\cite{BCL, {Ananth:2006fh}}
\bea
\label{lcg}
{\cal L}\;\;=&&\!\!\!\!\!\!\bh\, \square \, h+2\kappa\,\bh\,\parm^2\bigg(\frac{\bp}{\parm}h\frac{\bp}{\parm}h-h\frac{\bp^2}{\parm^2}h\bigg)+2\kappa\,h\,\parm^2\bigg(\frac{\p}{\parm}\bh\frac{\p}{\parm}\bh-\bh\frac{\p^2}{\parm^2}\bh\bigg) \nonumber \\
&&\!\!\!\!\!\!\!\!\, \nonumber \\
&&\!\!\!\!\!\!\!\!+2\kappa^2{\biggl \{}\,\fr{\parm^2}\big(\parm h\parm\bh\big)\frac{\p\bp}{\parm^2}\big(\parm h\parm\bh\big)+\fr{\parm^3}\big(\parm h\parm\bh\big)\left(\p\bp h\,\parm\bh+\parm h\p\bp\bh\right) \nonumber\\
&&\!\!\!\!\!\!\!\!-\fr{\parm^2}\big(\parm h\parm\bh\big)\,\left(2\,\p\bp h\,\bh+2\,h\p\bp\bh+9\,\bp h\p\bh+\p h\bp\bh-\frac{\p\bp}{\parm}h\,\parm\bh-\parm h\frac{\p\bp}{\parm}\bh\right) \nonumber  \\
&&\!\!\!\!\!\!\!\!-2\fr{\parm}\big(2\bp h\,\parm\bh+h\parm\bp\bh-\parm\bp h\bh\big)\,h\,\p\bh-2\fr{\parm}\big(2\parm h\,\p\bh+\parm\p h\,\bh-h\parm\p\bh\big)\,\bp h\,\bh \nonumber \\
&&\!\!\!\!\!\!\!\!-\fr{\parm}\big(2\bp h\,\parm\bh+h\parm\bp\bh-\parm\bp h\bh\big)\fr{\parm}\big(2\parm h\,\p\bh+\parm\p h\,\bh-h\parm\p\bh\big) \nonumber \\
&&\!\!\!\!\!\!\!\!-h\,\bh\,\bigg(\p\bp h\,\bh+h\p\bp\bh+2\,\bp h\p\bh+3\frac{\p\bp}{\parm}h\,\parm\bh+3\parm h\frac{\p\bp}{\parm}\bh\bigg){\biggr \}}\ .
\eea
 The terms containing a $\parp$ at order $\kappa^2$ were eliminated using a field redefinition~\cite{BCL}
\bea
\label{shift}
h\rightarrow h - \kappa^2\,\fr{\parm}{\biggl \{}2\,\parm^2h\fr{\parm^3}(\parm h\parm\bh)+\parm h\fr{\parm^2}(\parm h\parm\bh)+\fr{3}(h\bh\parm h-hh\parm\bh)\,{\biggr \}}\ .
\eea
\ndt Moving to five-point interaction vertices, ie. at order $\kappa^3$~\cite{Ananth:2008ik}, we find
\bea
{\cal L}_{\kappa^3}=2\sqrt{2}\kappa^3\,L_5\ ,
\eea
where $L_5$ reads 
\bea
\label{result}
L_5=&&\!\!\!\!\!\!-\fr{\sqrt 2}h\bp h\bp\bh\fr{\parm^2}(\parm\bh\parm h)+\frac{\sqrt 2}{3}\bh hh\bp h\bp\bh+\frac{\sqrt 2}{3}h\bp h\bp(\bh\bh h)+\frac{\sqrt 2}{3}h\bp\bh\bp(\bh hh) \nn \\
&&\!\!\!\!\!\!+\fr{\sqrt 2}h\bp h\bp\bh\fr{\parm^2}(\parm\bh\parm h)-\frac{\sqrt 2}{3}\bh hh\bp h\bp\bh-\frac{\sqrt 2}{3}h\bp h\bp(\bh\bh h)-\frac{\sqrt 2}{3}h\bp\bh\bp(\bh hh) \nn \\
&&\!\!\!\!\!\!-\frac{3}{4\sqrt 2}h\frac{\bp}{\parm^2}(\parm\bh\parm h)\frac{\bp}{\parm^2}(\parm\bh\parm h)+\fr{2\sqrt 2}h\fr{\parm^2}(\parm\bh\parm h)\frac{\bp\bp}{\parm^2}(\parm\bh\parm h) \nn \\
&&\!\!\!\!\!\!-\fr{3\sqrt 2}\bh hh\frac{\bp\bp}{\parm^2}(\parm\bh\parm h)-\fr{3\sqrt 2}h\frac{\bp\bp}{\parm^2}(\parm h\parm[\bh\bh h])-\fr{3\sqrt 2}h\frac{\bp\bp}{\parm^2}(\parm[\bh hh]\parm\bh)\nn \\
&&\!\!\!\!\!\!-\fr{2\sqrt 2}h\frac{\bp\bp}{\parm^2}(\parm[\bh h]\parm[\bh h])-2{\sqrt 2}\,\bh\bp h{\biggl [}\frac{\bp}{\parm}\{h\parm(\bh h)\}+\frac{\bp}{\parm}\{\fr{\parm^2}(\parm h\parm\bh)\parm h\} \nn \\
&&\!\!\!\!\!\!-\frac{\bp}{\parm}(h\bh\parm h)-\fr{3}\bp(hh\bh)+\fr{\parm}\{\fr{\parm^2}(\parm h\parm\bh)\bp\parm h\}\,{\biggr ]}+\fr{\sqrt 2}\,h\,\fr{\parm}{\biggl [}\parm h\bp\bh \nn \\
&&\!\!\!\!\!\!+\parm\bh\bp h-\frac{3}{2}\frac{\bp}{\parm}(\parm\bh\parm h)+2\bp(h\parm\bh)-\bp\parm(\bh h){\biggr ]}\,\times\,\fr{\parm}{\biggl [}\parm h\bp\bh+\parm\bh\bp h \nn \\
&&\!\!\!\!\!\!-\frac{3}{2}\frac{\bp}{\parm}(\parm\bh\parm h)+2\bp(h\parm\bh)-\bp\parm(\bh h){\biggr ]}+\bp h\bp h\,{\biggl [}\,\frac{\sqrt 2}{3}\bh\bh h+\frac{3}{\sqrt 2}\bh\fr{\parm^2}(\parm h\parm\bh)\,{\biggr ]} \nn \\
&&\!\!\!\!\!\!+{\sqrt 2}\bp h\,\fr{\parm}{\biggl \{}-\fr{\parm^2}(\parm h\parm\bh)\parm h\bp\bh+\fr{3}\parm h\bp(\bh\bh h)+\fr{3}\parm(\bh hh)\bp\bh+\fr{3}\parm\bh\bp(\bh hh) \nn \\
&&\!\!\!\!\!\!-\fr{\parm^2}(\parm h\parm\bh)\parm\bh\bp h+\fr{3}\parm(\bh\bh h)\bp h-\parm(\bh h)\bp(\bh h)-\fr{2}\frac{\bp}{\parm}[\parm h\parm(\bh\bh h)] \nn \\
&&\!\!\!\!\!\!+\frac{3}{2}\fr{\parm^2}(\parm h\parm\bh)\frac{\bp}{\parm}(\parm\bh\parm h)-\fr{2}\frac{\bp}{\parm}[\parm(\bh hh)\parm\bh]+\frac{3}{4}\frac{\bp}{\parm}[\parm(\bh h)\parm(\bh h)] \nn \\
&&\!\!\!\!\!\!-\fr{2}\frac{\bp}{\parm^2}(\parm\bh\parm h)\fr{\parm}(\parm\bh\parm h)+\frac{2}{3}\bp[h\parm(\bh\bh h)]+\frac{2}{3}\bp[\bh hh\parm\bh]-\fr{6}\bp\parm(\bh\bh hh) \nn \\
&&\!\!\!\!\!\!-2\bp[\fr{\parm^2}(\parm h\parm\bh)h\parm\bh]-\bp[\bh h\parm(\bh h)]+\bp[\fr{\parm^2}(\parm\bh\parm h)\parm(\bh h)]\,{\biggr \}} \nn \\
&&\!\!\!\!\!\!-{\sqrt 2}\fr{\parm}{\biggl \{}\fr{\parm^2}(\parm h\parm\bh)\bp\parm h-\fr{3}\bp\parm(hh\bh)-\bp(h\bh\parm h)+\bp[\fr{\parm^2}(\parm h\parm\bh)\parm h] \nn \\
&&\!\!\!\!\!\!+\bp(h\parm(\bh h){\biggr \}}\times\,\fr{\parm}{\biggl \{}\parm h\bp\bh+\parm\bh\bp h-\frac{3}{2}\frac{\bp}{\parm}(\parm\bh\parm h)+2\bp(h\parm\bh)-\bp\parm(\bh h)\,{\biggr \}} \nn \\
\, \nn \\
&&\!\!\!\!\!\!\!\!\!\!\!\!+{\sqrt 2}[\bh h+\frac{3}{2}\fr{\parm^2}(\parm h\parm\bh)]\,\fr{\parm}{\biggl \{}\parm h\bp\bh+\parm\bh\bp h-\frac{3}{2}\frac{\bp}{\parm}(\parm\bp\parm h)+2\bp(h\parm\bh)-\bp\parm(\bh h)\,{\biggr \}}\bp h \nn \\
&&\!\!\!\!\!\!\!\!\!\!\!\!+{\biggl \{}2\,\parm^2\bh\fr{\parm^3}(\parm h\parm\bh)+\parm\bh\fr{\parm^2}(\parm h\parm\bh)+\fr{3}(h\bh\parm \bh-\bh\bh\parm h)\,{\biggr \}}\parm\bigg(\frac{\bp}{\parm}h\frac{\bp}{\parm}h-h\frac{\bp^2}{\parm^2}h\bigg) \nn \\
&&\!\!\!\!\!\!\!\!\!\!\!\!+{\biggl \{}2\,\parm^2h\fr{\parm^3}(\parm h\parm\bh)+\parm h\fr{\parm^2}(\parm h\parm\bh)+\fr{3}(h\bh\parm h-hh\parm\bh)\,{\biggr \}}\times{\biggl [}2\frac{\bp}{\parm^2}(\parm^2\bh\frac{\bp}{\parm}h) \nn \\
&&\!\!\!\!\!\!\!\!\!\!\!\!+\parm(\frac{\bp}{\parm}h\frac{\bp}{\parm}h-h\frac{\bp^2}{\parm^2}h)-\fr{\parm}(\parm^2\bh\frac{\bp^2}{\parm^2}h)-\frac{\bp^2}{\parm^3}(h\parm^2\bh){\biggr ]}\,\,+\,{\mbox {c.c.}}
\eea
\vskip 0.3cm
\ndt The last three lines of (\ref {result}) represent quintic interaction vertices produced by applying the field redefinition in (\ref {shift}) to the cubic interaction vertices in (\ref {lcg}). Although $L_5$ appears to contain non-MHV structures, these are related by conjugation to MHV vertices.

\vskip 0.5cm
\section{The 6-point result}

\ndt One new result in this paper is the perturbative expansion to order $\kappa^4$, from the closed form action (\ref {aaction}). This (six) is the lowest order at which truly non-MHV structures first appear. We find
\bea
{\cal L}_{\kappa^4}=4\kappa^4\,L_6\ ,
\eea
where
\bea
\label{finresult}
&&\!\!\!\!\!\!\!\!\!\! L_6=4 \psi_{4} \partial\bar{\partial} \psi_{2}+\psi_{2}^{2} \partial\bar{\partial} \psi_{2}-\frac{1}{15}\left(h {\partial \bar{\partial}}[(h \bar{h})^2\bar{h}]+\bar{h} {\partial \bar{\partial}}[(h \bar{h})^2h]\right)+\frac{1}{3} h \bar{h}{\partial \bar{\partial}}[(h \bar{h})^2] \nn \\
&&\!\!\!\!\!\!\!\!\!\!-\frac{2}{9} h \bar{h} h{\partial \bar{\partial}}[h \bar{h} \bar{h}]+\frac{\psi_{2}}{3}\left(\frac{\partial \bar{\partial}}{\partial_-} h \partial_{-}[h \bar{h} \bar{h}]+\partial_{-} \bar{h} \frac{\partial \bar{\partial}}{\partial_-}[h \bar{h} h]+\partial_{-} h \frac{\partial \bar{\partial}}{\partial_-}[h \bar{h} \bar{h}]+\frac{\partial \bar{\partial}}{\partial_-} \bar{h} \partial_{-}[h \bar{h} h]\right) \nn \\
&&\!\!\!\!\!\!\!\!\!\!-\psi_{2} \frac{\partial \bar{\partial}}{\partial_-}[h \bar{h}] \partial_{-}[h \bar{h}]+\left(\psi_{4}+\frac{\psi_{2}^{2}}{2}\right)\left(\frac{\partial \bar{\partial}}{\partial_-} h \partial_{-} \bar{h}+\partial_{-} h \frac{\partial \bar{\partial}}{\partial_-} \bar{h}\right)-\frac{1}{6}(h \bar{h} h \bar{h}) \partial \bar{\partial} \psi_{2}-h \bar{h} \partial \bar{\partial} \psi_{4}\nn \\
&&\!\!\!\!\!\!\!\!\!\!-\frac{\psi_{2}}{2} h \bar{h} \partial \bar{\partial} \psi_{2}+\frac{3}{4} h \bar{h} \partial \psi_{2} \bar{\partial} \psi_{2}-\frac{5}{2} \psi_{4} \partial \bar{\partial} \psi_{2}-\frac{5}{16} \psi_{2}^{2} \partial \bar{\partial} \psi_{2}\nn \\
&&\!\!\!\!\!\!\!\!\!\!+(\partial h \bar{\partial} \bar{h}-\partial \bar{h} \bar{\partial} h)\left(\frac{1}{6} h \bar{h} h \bar{h}+\frac{1}{2} \psi_{2} h \bar{h}+\frac{1}{2} \psi_{4}+\frac{1}{8} \psi_{2}^{2}\right)\nn \\
&&\!\!\!\!\!\!\!\!\!\!+\left(\frac{1}{3} h \bar{h}+\frac{1}{6} \psi_{2}\right)(\partial h \bar{\partial}[h \bar{h} \bar{h}]+\bar{\partial} \bar{h} \partial[h \bar{h} h]-\bar{\partial} h \partial[h \bar{h} \bar{h}]-\partial \bar{h} \bar{\partial}[h \bar{h} h])\nn \\
&&\!\!\!\!\!\!\!\!\!\!+\left(\frac{h \bar{h}}{3}+\frac{\psi_{2}}{2}\right)(h \partial \bar{h}-\bar{h} \partial h) \bar{\partial}[h \bar{h}]+\left(\frac{h \bar{h}}{3}+\frac{\psi_{2}}{2}\right)(\bar{h} \bar{\partial} h-h \bar{\partial} \bar{h}) \partial[h \bar{h}]+\frac{1}{6}(h \partial \bar{h}-\bar{h} \partial h) \bar{\partial}[h \bar{h} h \bar{h}]\nn \\
&&\!\!\!\!\!\!\!\!\!\!+\frac{1}{6}(\bar{h} \bar{\partial} h-h \bar{\partial} \bar{h}) \partial[h \bar{h} h \bar{h}]+\frac{h}{3}(\bar{\partial}[h \bar{h}] \partial[h \bar{h} \bar{h}]-\bar{\partial}[h \bar{h} \bar{h}] \partial[h \bar{h}])+\frac{\bar{h}}{3}(\partial[h \bar{h}] \bar{\partial}[h \bar{h} h]-\partial[h \bar{h} h] \bar{\partial}[h \bar{h}])\nn \\
&&\!\!\!\!\!\!\!\!\!\!+\biggl\{2 h \partial \bar{h} \frac{\bar{\partial}}{\partial_{-}}\left[\frac{2}{3} h \bar{h} \bar{h} \partial_{-} h-\frac{2}{3} h \bar{h} h \partial_{-} \bar{h}-2 \psi_{2} h \partial_{-} \bar{h}+\psi_{2} \partial_{-}[h \bar{h}]\right]-2 h \partial \bar{h} \frac{1}{\partial_{-}}\biggl[\frac {1}{3} (\partial_{-} h \bar{\partial}[h \bar{h} \bar{h}]\nn \\
&&\!\!\!\!\!\!\!\!\!\!+\bar{\partial} \bar{h} \partial_{-}[h \bar{h} h]+\bar{\partial} h \partial_{-}[h \bar{h} \bar{h}]+\partial_{-} \bar{h} \bar{\partial}[h \bar{h} h])-\partial_{-}[h \bar{h}] \bar{\partial}[h \bar{h}]+\psi_{2}\left(\partial_{-} h \bar{\partial} \bar{h}+\bar{\partial} h \partial_{-} \bar{h}\right) \nn \\
&&\!\!\!\!\!\!\!\!\!\!+\frac{3}{2} \psi_{2} \partial_{-} \bar{\partial} \psi_{2}-\frac{1}{2} \bar{\partial} \psi_{2} \partial_{-} \psi_{2}+\frac{3}{2} \partial_{-} \bar{\partial} \psi_{4}\biggl]+ \left(\bar{\partial}\left[h \bar{h}-2 \frac{1}{\partial_{-}}\left[h \partial_{-} \bar{h}\right]\right]-\frac{1}{\partial_{-}}\left[\partial_{-} h \bar{\partial} \bar{h}+\bar{\partial} h \partial_{-} \bar{h}\right]-\frac{3}{2} \bar{\partial} \psi_{2}\right) \nn \\
&&\!\!\!\!\!\!\!\!\!\! \times\left(2 h \partial\left[\frac{1}{3} h \bar{h} \bar{h}+\frac{1}{\partial_{-}}\left[\psi_{2} \partial_{-} \bar{h}-\bar{h} \bar{h} \partial_{-} h\right]\right]+2 \partial \bar{h}\left[\frac{1}{3} h \bar{h} h-\frac{3}{2} \psi_{2} h\right]\right) + c.c. \biggl\}\nn \\
&&\!\!\!\!\!\!\!\!\!\!-\biggl(\frac{1}{3}(h \bar{h})^{2}-3 \psi_{4}-3 \psi_{2} h \bar{h}+\frac{9}{4} \psi_{2}^{2}\biggl) \bar{\partial} h \partial \bar{h}\nn \\&&\!\!\!\!\!\!\!\!\!\!-\left(h \bar{h}-\frac{3}{2} \psi_{2}\right)\left(2 \bar{\partial} h \partial\left[\frac{1}{3} h \bar{h} \bar{h}+\frac{1}{\partial_{-}}\left[\psi_{2} \partial_{-} \bar{h}-\bar{h} \bar{h} \partial_{-} h\right]\right]+\text { c.c. }\right)\nn \\
&&\!\!\!\!\!\!\!\!\!\!-\left(h \bar{h}-\frac{3}{2} \psi_{2}\right)\left\{\bar{\partial}\left[h \bar{h}-2 \frac{1}{\partial_{-}}\left[h \partial_{-} \bar{h}\right]\right]-\frac{1}{\partial_{-}}\left[\partial_{-} h \bar{\partial} \bar{h}+\bar{\partial} h \partial_{-} \bar{h}\right]-\frac{3}{2} \bar{\partial} \psi_{2}\right\}\left\{\partial\left[h \bar{h}-2 \frac{1}{\partial_{-}}\left[\bar{h} \partial_{-} h\right]\right]\right. \nn\\ &&\!\!\!\!\!\!\!\!\!\!\left.-\frac{1}{\partial_{-}}\left[\partial_{-} \bar{h} \partial h+\partial \bar{h} \partial_{-} h\right]-\frac{3}{2} \partial \psi_{2}\right\} \nn \\
&&\!\!\!\!\!\!\!\!\!\! -\left\{2 \bar{\partial} h \frac{\partial}{\partial_{-}}\left(-\frac{2}{15} \bar{h} \partial_{-}[h \bar{h}]^{2}-\frac{2}{3} \psi_{2} \bar{h} \partial_{-}[h \bar{h}]+\left(\frac{8}{15}(h \bar{h})^{2}+\frac{4}{3} \psi_{2} h \bar{h}+\psi_{4}+\frac{1}{2} \psi_{2}^{2}\right) \partial_{-} \bar{h}\right)+\text { c.c. }\right\}
\nn \\
&&\!\!\!\!\!\!\!\!\!\!-\left\{\left( \partial \left[ h \bar { h } - 2 \frac { 1 } { \partial _ { - } } [ \bar { h } \partial _ { - } h ] \right] - \frac { 1 } { \partial _ { - } } [ \partial _ { - } \bar { h } \partial h + \partial \bar { h } \partial _ { - } h ] - \frac { 3 } { 2 } \partial \psi _ { 2 } \right)\right.\nn \\
&&\!\!\!\!\!\!\!\!\!\!\times\biggl(\frac { \bar { \partial } } { \partial _ { - } } \left[\frac{2}{3} h \bar{h} \bar{h} \partial_{-} h-\frac{2}{3} h \bar{h} h \partial_{-} \bar{h} -2 \psi_{2} h \partial_{-} \bar{h}+\psi_{2} \partial_{-}[h \bar{h}]\right]\nn \\
&&\!\!\!\!\!\!\!\!\!\!-\frac{1}{\partial_{-}}\biggl[\frac{1}{3}\left(\partial_{-} h \bar{\partial}[h \bar{h} \bar{h}]+\bar{\partial} \bar{h} \partial_{-}[h \bar{h} h]+\bar{\partial} h \partial_{-}[h \bar{h} \bar{h}]+\partial_{-} \bar{h} \bar{\partial}[h \bar{h} h]\right)-\partial_{-}[h \bar{h}] \bar{\partial}[h \bar{h}]+\psi_{2}\left(\partial_{-} h \bar{\partial} \bar{h}+\bar{\partial} h \partial_{-} \bar{h}\right)\nn \\
&&\!\!\!\!\!\!\!\!\!\!+\frac{3}{2} \psi_{2} \partial_{-} \bar{\partial} \psi_{2}-\frac{1}{2} \bar{\partial} \psi_{2} \partial_{-} \psi_{2}+\frac{3}{2} \partial_{-} \bar{\partial} \psi_{4}\biggl]\biggl)+\text { c.c. }\biggl\}
\nn \\
&&\!\!\!\!\!\!\!\!\!\!-2 \partial\left[\frac{1}{3} h \bar{h} \bar{h}+\frac{1}{\partial_{-}}\left[\psi_{2} \partial_{-} \bar{h}-\bar{h} \bar{h} \partial_{-} h\right]\right] \bar{\partial}\left[\frac{1}{3} h h \bar{h}+\frac{1}{\partial_{-}}\left[\psi_{2} \partial_{-} h-h h \partial_{-} \bar{h}\right]\right]\nn \\
&&\!\!\!\!\!\!\!\!\!\!+\biggl\{\frac{\partial \bar{\partial}}{\partial_-} \bar{M}\left[\partial_{-}^{2} h \frac{1}{\partial_{-}^{3}}\left(\partial_{-} h \partial_{-} \bar{h}\right)+\frac{1}{3}\left(h \bar{h} \partial_{-} h-h h \partial_{-} \bar{h}\right)\right]-\frac{1}{\partial_{-}^2}\left(\partial_{-} M \partial_{-} \bar{h}\right)\biggl[-\frac{1}{2}\frac{\partial \bar{\partial}}{\partial_-^2}\left(\partial_{-} h \partial_{-} \bar{h}\right)\nn \\
&&\!\!\!\!\!\!\!\!\!\!+\frac{1}{\partial_{-}}\left(\partial \bar{\partial} h \partial_{-} \bar{h}\right)\biggl]- M\bar{h}\left[ \bar{\partial}h \partial \bar{h}+\frac{4}{3} h \partial \bar{\partial} \bar{h}+4 \partial_{-} h \frac{\partial \bar{\partial}}{\partial_-} \bar{h}\right]-2 \frac{1}{\partial_{-}^{2}}\left(\partial_{-} M \partial_{-} \bar{h}\right)\biggl[\frac{1}{2}\bar{\partial} h \partial \bar{h}+h \partial\bar{\partial} \bar{h}-\partial_{-} h \frac{\partial \bar{\partial}}{\partial_-} \bar{h}\nn \\
&&\!\!\!\!\!\!\!\!\!\!+\frac{1}{\partial_{-}}\left(\partial \bar{\partial} h \partial_{-} \bar{h}\right)-\frac{1}{4} \frac{\partial \bar{\partial}}{\partial_{-}^{2}}\left(\partial_{-} h \partial_{-} \bar{h}\right)\biggl]+2 M \partial \bar{h} \frac{1}{\partial_{-}}\left(2 \bar{\partial} h \partial_{-} \bar{h}+h \partial_{-} \bar{\partial} \bar{h}-\partial_{-} \bar{\partial} h \bar{h})+ c.c \right. \biggl\}\nn \\
&&\!\!\!\!\!\!\!\!\!\!-\frac{1}{\partial_{-}}\left(2 \bar{\partial} M \partial_{-} \bar{h}+M \partial_{-} \bar{\partial}\bar{h}-\partial_{-} \bar{\partial}M \bar{h}\right) \frac{1}{\partial_{-}}\left(2 \partial_{-} h \partial \bar{h}+\partial_{-} \partial h \bar{h}-h \partial_{-} \partial \bar{h}\right) \nn \\
&&\!\!\!\!\!\!\!\!\!\!+\biggl\{\frac{\partial \bar{\partial}}{\partial_-} \bar{h}\left[\partial_{-}^{2} M \frac{1}{\partial_{-}^{3}}\left(\partial_{-} h \partial_{-} \bar{h}\right)+\frac{1}{3}\left(M \bar{h} \partial_{-} h-M h \partial_{-} \bar{h}\right)\right]-\frac{1}{\partial_{-}^2}\left(\partial_{-} h \partial_{-} \bar{M}\right)\biggl[-\frac{1}{2}\frac{\partial \bar{\partial}}{\partial_-^2}\left(\partial_{-} h \partial_{-} \bar{h}\right)\nn \\
&&\!\!\!\!\!\!\!\!\!\!+\frac{1}{\partial_{-}}\left(\partial \bar{\partial} h \partial_{-} \bar{h}\right)\biggl]- h\bar{M}\left[ \bar{\partial}h \partial \bar{h}+\frac{4}{3} h \partial \bar{\partial} \bar{h}+4 \partial_{-} h \frac{\partial \bar{\partial}}{\partial_-} \bar{h}\right]-2 \frac{1}{\partial_{-}^{2}}\left(\partial_{-} h \partial_{-} \bar{M}\right)\biggl[\frac{1}{2}\bar{\partial} h \partial \bar{h}+h \partial\bar{\partial} \bar{h}-\partial_{-} h \frac{\partial \bar{\partial}}{\partial_-} \bar{h}\nn \\
&&\!\!\!\!\!\!\!\!\!\!+\frac{1}{\partial_{-}}\left(\partial \bar{\partial} h \partial_{-} \bar{h}\right)-\frac{1}{4} \frac{\partial \bar{\partial}}{\partial_{-}^{2}}\left(\partial_{-} h \partial_{-} \bar{h}\right)\biggl]+2 h \partial \bar{M} \frac{1}{\partial_{-}}\left(2 \bar{\partial} h \partial_{-} \bar{h}+h \partial_{-} \bar{\partial} \bar{h}-\partial_{-} \bar{\partial} h \bar{h})+ c.c \right.\biggl\}\nn \\
&&\!\!\!\!\!\!\!\!\!\!-\frac{1}{\partial_{-}}\left(2 \bar{\partial} h \partial_{-} \bar{M}+h \partial_{-} \bar{\partial}\bar{M}-\partial_{-} \bar{\partial}h \bar{M}\right) \frac{1}{\partial_{-}}\left(2 \partial_{-} h \partial \bar{h}+\partial_{-} \partial h \bar{h}-h \partial_{-} \partial \bar{h}\right)\nn \\
&&\!\!\!\!\!\!\!\!\!\!+\biggl\{\frac{\partial \bar{\partial}}{\partial_-} \bar{h}\left[\partial_{-}^{2} h \frac{1}{\partial_{-}^{3}}\left(\partial_{-} M \partial_{-} \bar{h}\right)+\frac{1}{3}\left(h \bar{M} \partial_{-} h-h M \partial_{-} \bar{h}\right)\right]-\frac{1}{\partial_{-}^2}\left(\partial_{-} h \partial_{-} \bar{h}\right)\biggl[-\frac{1}{2}\frac{\partial \bar{\partial}}{\partial_-^2}\left(\partial_{-} M \partial_{-} \bar{h}\right)\nn \\
&&\!\!\!\!\!\!\!\!\!\!+\frac{1}{\partial_{-}}\left(\partial \bar{\partial} M \partial_{-} \bar{h}\right)\biggl]- h\bar{h}\left[ \bar{\partial}M \partial \bar{h}+\frac{4}{3} M \partial \bar{\partial} \bar{h}+4 \partial_{-} M \frac{\partial \bar{\partial}}{\partial_-} \bar{h}\right]-2 \frac{1}{\partial_{-}^{2}}\left(\partial_{-} h \partial_{-} \bar{h}\right)\biggl[\frac{1}{2}\bar{\partial} M \partial \bar{h}+M \partial\bar{\partial} \bar{h}\nn \\
&&\!\!\!\!\!\!\!\!\!\!-\partial_{-} M \frac{\partial \bar{\partial}}{\partial_-} \bar{h}
+\frac{1}{\partial_{-}}\left(\partial \bar{\partial} M \partial_{-} \bar{h}\right)-\frac{1}{4} \frac{\partial \bar{\partial}}{\partial_{-}^{2}}\left(\partial_{-} M \partial_{-} \bar{h}\right)\biggl]+2 h \partial \bar{h} \frac{1}{\partial_{-}}\left(2 \bar{\partial} M \partial_{-} \bar{h}+M \partial_{-} \bar{\partial} \bar{h}-\partial_{-} \bar{\partial} M \bar{h})+ c.c \right.\biggl\}\nn \\
&&\!\!\!\!\!\!\!\!\!\!-\frac{1}{\partial_{-}}\left(2 \bar{\partial} h \partial_{-} \bar{h}+h \partial_{-} \bar{\partial}\bar{h}-\partial_{-} \bar{\partial}h \bar{h}\right) \frac{1}{\partial_{-}}\left(2 \partial_{-} M \partial \bar{h}+\partial_{-} \partial M \bar{h}-M \partial_{-} \partial \bar{h}\right)\nn \\
&&\!\!\!\!\!\!\!\!\!\!+\biggl\{\frac{\partial \bar{\partial}}{\partial_-} \bar{h}\left[\partial_{-}^{2} h \frac{1}{\partial_{-}^{3}}\left(\partial_{-} h \partial_{-} \bar{M}\right)+\frac{1}{3}\left(h \bar{h} \partial_{-} M-h h \partial_{-} \bar{M}\right)\right]-\frac{1}{\partial_{-}^2}\left(\partial_{-} h \partial_{-} \bar{h}\right)\biggl[-\frac{1}{2}\frac{\partial \bar{\partial}}{\partial_-^2}\left(\partial_{-} h \partial_{-} \bar{M}\right)\nn \\
&&\!\!\!\!\!\!\!\!\!\!+\frac{1}{\partial_{-}}\left(\partial \bar{\partial} h \partial_{-} \bar{M}\right)\biggl]- h\bar{h}\left[ \bar{\partial}h \partial \bar{M}+\frac{4}{3} h \partial \bar{\partial} \bar{M}+4 \partial_{-} h \frac{\partial \bar{\partial}}{\partial_-} \bar{M}\right]-2 \frac{1}{\partial_{-}^{2}}\left(\partial_{-} h \partial_{-} \bar{h}\right)\biggl[\frac{1}{2}\bar{\partial} h \partial \bar{M}+h \partial\bar{\partial} \bar{M}\nn \\
&&\!\!\!\!\!\!\!\!\!\!-\partial_{-} h \frac{\partial \bar{\partial}}{\partial_-} \bar{M}
+\frac{1}{\partial_{-}}\left(\partial \bar{\partial} h \partial_{-} \bar{M}\right)-\frac{1}{4} \frac{\partial \bar{\partial}}{\partial_{-}^{2}}\left(\partial_{-} h \partial_{-} \bar{M}\right)\biggl]+2 h \partial \bar{h} \frac{1}{\partial_{-}}\left(2 \bar{\partial} h \partial_{-} \bar{M}+h \partial_{-} \bar{\partial} \bar{M}-\partial_{-} \bar{\partial} h \bar{M})+ c.c \right.\biggl\}\nn \\
&&\!\!\!\!\!\!\!\!\!\!-\frac{1}{\partial_{-}}\left(2 \bar{\partial} h \partial_{-} \bar{h}+h \partial_{-} \bar{\partial}\bar{h}-\partial_{-} \bar{\partial}h \bar{h}\right) \frac{1}{\partial_{-}}\left(2 \partial_{-} h \partial \bar{M}+\partial_{-} \partial h\bar{M}-h \partial_{-} \partial \bar{M}\right)\,\ ,
\eea
with
\bea
M = -\frac{1}{2}\,\fr{\parm}{\biggl \{}2\,\parm^2h\fr{\parm^3}(\parm h\parm\bh)+\parm h\fr{\parm^2}(\parm h\parm\bh)+\fr{3}(h\bh\parm h-hh\parm\bh)\,{\biggr \}}\ .
\eea
\ndt The last 16 lines of (\ref {finresult}) arise as a consequence of applying the field redefinition in (\ref {shift}) on the quartic interaction vertices in (\ref {lcg}). As done earlier at order $\kappa^2$, at this order also, interaction vertices containing a $\parp$ (including ones resulting from (\ref{shift})), need to be eliminated. This is achieved by the following field redefinition
\bea
\label{shift2}
h &&\!\!\!\!\!\!\!\!\rightarrow h + 2\kappa^4\biggl\{\frac{1}{\partial_-}\biggl[\frac{2}{3}h\partial_-\bar{h}M - \frac{5}{6}\partial_-h\bar{h}M - \frac{1}{2}h\bar{h}\partial_-M + \frac{1}{2}\partial_-\bar{M}h h + \frac{1}{6}\bar{M}h\partial_-h\nn\\&&
 -\partial_-^2h\frac{1}{\partial_-^3}\left(\partial_-{M}\partial_-\bar{h} \right) -\partial_-^2M\frac{1}{\partial_-^3}\left(\partial_-{h}\partial_-\bar{h} \right) -\partial_-^2h\frac{1}{\partial_-^3}\left(\partial_-\bar{M}\partial_-{h} \right) - \partial_-h\left(\psi_4 + \frac{\psi_2^2}{2}\right) +\frac{1}{3}h\partial_-[h\bar{h}]^2 \nn\\&&
  + h\psi_2\partial_-[h\bar{h}] - \frac{h^2}{3}\left(\frac{1}{3}\partial_-[h\bar{h}\bar{h}] + 2\psi_2\partial_-\bar{h}\right) -\frac{2h \bar{h}}{3}\left(\frac{1}{3}\partial_-[h\bar{h}h] + 2\psi_2\partial_-h\right) \biggl]\nn\\&&
 -2 \partial_-h\frac{1}{\partial_-^3}(\partial_-M\partial_-\bar{h}) -2 \partial_-h\frac{1}{\partial_-^3}(\partial_-h\partial_-\bar{M}) +\partial_-h\frac{1}{\partial_-}(4\psi_4 + \psi_2^2) - \frac{1}{15}(h\bar{h})^2h \nn\\&&
-\frac{1}{2}M\psi_2 + \frac{1}{2}\partial_-h\frac{1}{\partial_-^2}(M\partial_-\bar{h}) + \frac{1}{2}\partial_-h\frac{1}{\partial_-^2}(\bar{M}\partial_-{h})  \biggl\}\ .
\eea

\ndt Since the $\parp$ that occurs in (\ref{aaction}) appears linearly, it can always be shifted to a higher order.

\vskip 0.5cm

\section{Spinor helicity and amplitudes}
\ndt Now that we have perturbative expansions of the closed form action, we can examine these vertices in momentum space. Before doing so, we briefly review the intimate link between light-cone gauge and spinor helicity variables~\cite{Ananth:2012un}. A four-vector $p_\mu$ may be rewritten as a matrix $p_{a\dot{a}}$ using the Pauli matrices $\sigma_{\mu}=(-\mathbf{1},\sigma)$
\bea
\label{bispinor}
p_{\mu}\left(\sigma^{\mu}\right)_{a \dot{a}} \equiv p_{a \dot{a}} =\left(\begin{array}{cc}
-p_{0}+p_{3} & p_{1}-i p_{2} \\
p_{1}+i p_{2} & -p_{0}-p_{3}
\end{array}\right)=\sqrt{2}\left(\begin{array}{cc}
-p_{-} & \bar{p} \\
p & -p_{+}
\end{array}\right)\ .
\eea
The determinant of this matrix is the length of the four-vector $p_\mu$
\bea
\operatorname{det}\left(p_{a \dot{a}}\right)=-2\left(p \bar{p}-p_{+} p_{-}\right)=-p^{\mu} p_{\mu} .
\eea
The on-shell condition for a light-like vector $p_{\mu}$ is $p_{+}=\frac{p \bar{p}}{p_{-}}$. We define holomorphic and anti-holomorphic spinors by
\bea
\lambda_{a}=\frac{2^{\frac{1}{4}}}{\sqrt{p}_{-}}\left(\begin{array}{c}
p_{-} \\
-p
\end{array}\right), \quad \tilde{\lambda}_{\dot{a}}=-\left(\lambda_{a}\right)^{*}=-\frac{2^{\frac{1}{4}}}{\sqrt{p}_{-}}\left(\begin{array}{c}
p_{-} \\
-\bar{p}
\end{array}\right),
\eea
such that $\lambda_{a} \tilde{\lambda}_{\dot{a}}$ reproduces (\ref{bispinor}) on-shell. We define the off-shell holomorphic and anti-holomorphic spinor products~\cite{Brandhuber},
\bea
\langle i j\rangle=\sqrt{2}\, \frac{p^{i}\, p_{-}^{j}\,-\,p^{j}\, p_{-}^{i}}{\sqrt{p_{-}^{i}\, p_{-}^{j}}}, \quad[i j]=\sqrt{2}\, \frac{\bar{p}\,^{i}\, p_{-}^{j}\,-\,\bar{p}\,^{j}\, p_{-}^{i}}{\sqrt{p_{-}^{i}\, p_{-}^{j}}} .
\eea

\vskip 0.3cm
\subsection{MHV Lagrangians}
\vskip 0.2cm
In the light-cone gauge, the Lagrangian for Yang-Mills theory has the following helicity structure
\bea
\label{ym}
{\it L}_{\mbox {YM}}={\it L}_{+-}\;+\;{\it L}_{++-}\;+\;{\it L}_{+--}\;+\;{\it L}_{++--}\ .
\eea
Feynman diagrams primarily constructed using the first cubic vertex vanish since these correspond to amplitudes involving all `plus' or all but one `plus' helicities~\cite{Grisaru,Elvang}.  This motivates the need for a canonical field redefinition, one that maps the kinetic term and the ${\it L}_{++-}$ (non-MHV) cubic vertex to a purely kinetic term. In momentum space, we seek a transformation of the form
\bea
\label{ymc}
{\it L}_{+\,-}\,+\,{\it L}_{+\,+\,-}\,\rightarrow\,{\it L'}_{+\,-}\ .
\eea
\ndt This is achieved through field redefinitions of the form~\cite{Mansf,Gorsky}
\bea
\label{ymr}
A\,&\sim\,&u_1\, B\,+\,u_2\, BB\,+\,u_3\, BBB\,+....\ ,\nn\\
\bar{A}\,&\sim\,&v_1\,\widetilde{B}\,+\,v_2\,\widetilde{B}B\,+\,v_3\,\widetilde{B}BB\,+....\ ,
\eea
\ndt where $A,\bar{A}$ are $+$ and $-$ helicity Yang-Mills fields and $B,\widetilde{B}$ are the shifted fields. The relations (\ref{ymr}) are purely schematic and we refer the reader to~\cite{Mansf,Gorsky} for details. The coefficients $u_n,v_n$ are functions of momenta and can be worked out, order by order, based on the requirement in (\ref{ymc}). In this process of eliminating the cubic non-MHV vertex from (\ref{ym}), the terms ${\it L}_{+--}$ and ${\it L}_{++--}$ generate an infinite series of MHV vertices. This new `MHV Lagrangian'~\cite{Ananth:2011hu} has the following structure
\bea
{\it L}_{\mbox {YM}}={\it L'}_{+-}\;+\;{\it L'}_{+--}\;+\;{\it L'}_{++--}\;+\;{\it L'}_{+++--}\;+\;{\it L'}_{++++--}\;+\;\ldots\;\;\;\;.
\eea
The key advantage of this Lagrangian is that all MHV amplitudes are trivial to read off (by simply taking the relevant momenta on-shell).
\vskip 0.3cm
\ndt The light-cone action for pure gravity is similar in helicity structure to Yang-Mills theory~\cite{Ananth:2010uy,Ananth:2017xpj}. The same procedure can be followed to eliminate the non-MHV cubic vertex from (\ref {lcg}). The canonical field redefinitions for $h,\bar{h}$ in momentum space take the form~\cite{Ananth:2007zy} 
\bea
\label{redef}
h\, &\sim\,& y_1\, C\,+\,y_2\, CC\,+\,y_3\, CCC\,+....\ ,\nn \\
\bar{h}\, &\sim\,& z_1\,\widetilde{C}\,+\,z_2\,\widetilde{C}C\,+\,z_3\,\widetilde{C}CC\,+....\ ,
\eea
where $C,\widetilde{C}$ are the shifted fields and the coefficients $y_n,z_n$ are functions of momenta. Once these field redefinitions are applied to the gravity Lagrangian, tree-level (MHV) amplitudes are trivial to read off. For example, the cubic MHV term in (\ref{lcg}) (in momentum space) reads
\bea
M_3(+,-,-)=\frac{{\langle k\,l\rangle}^6}{{\langle l\,p\rangle}^2{\langle p\,k\rangle }^2}\ ,
\eea
where $p$ is the momentum of the positive helicity field and $k,l$ the momenta of the negative helicity fields. At quartic order, these steps produce
\bea
M_4(+,+,-,-)= 
\frac{\langle k\,l\rangle^{8}[k\,l]}{\langle k\,l\rangle\langle k\,p\rangle\langle k\,q\rangle\langle l\,p\rangle\langle l\,q\rangle\langle p\, q\rangle^{2}}\ ,
\eea
where $p,q$ are the momenta of the positive helicity fields and $k,l$ the momenta of negative helicity fields. These off-shell MHV vertices factorise, as expected from the KLT relations~\cite{klt}, into products of off-shell MHV vertices in Yang-Mills theory~\cite{Ananth:2010uy}.

\vskip 0.5cm

\ndt \textbf{MHV structures at order $\kappa^4$}
\vskip 0.5cm
\ndt We note that (\ref {finresult}) does not contain structures like $\bh \bh hhhh$, the MHV vertex at order $\kappa^4$. In other words, MHV vertices do not arise naturally at this order and beyond. Instead, the field redefinition described above will produce such vertices. Alternatively, one could arrive at the $(++++--)$ amplitude by summing over the various Feynman diagrams, at order $\kappa^4$, arising from lower-point vertices. Building a MHV Lagrangian~\cite{Ananth:2011hu} to order $\kappa^4$, would involve working out the coefficients in (\ref{redef}) up to $n=4$.
\vskip 0.3cm

\section{Higher-order interaction vertices in light-cone gravity}
\vskip 0.3cm
\ndt Poincar\'e invariance needs to be explicitly verified in light-cone field theories. This check can be used as a tool to impose constraints on field theories in the light-cone gauge. Essentially, this means that light-cone consistency checks serve as a first-principles approach to deriving Lagrangians for interacting field theories. Specifically, the procedure involves starting with a general ansatz for interaction vertices and allowing the Poincar\'e  algebra to constrain the ansatz.
\vskip 0.3cm
\ndt  A massless field of integer spin $\lambda$, in light-cone gauge, has two physical degrees of freedom $\phi$ and $\bar{\phi}$ corresponding to the $+$ and $-$ helicity states respectively. The generators of the Poincar\'{e} algebra are the momenta
\bea
P^-=-i\frac{\partial \bar\partial}{\partial^{+}}=-P_+\ , \qquad P^+=-i\partial^+=-P_-\ , \qquad P=-i\partial\ , \qquad \bar P=-i\bar\partial\ ,
\eea
the rotation generators
\begin{eqnarray}
J = (x \bar\partial -\bar x \partial - \lambda) \;,\qquad  J^+ = i(x\partial^{+}-x^+\partial)\ , \nn\\
J^{+-}=i(x^-\partial^{+}-x^+\frac{\partial \bar\partial}{\partial^{+}}) \;,\qquad J^{-}=i(x\frac{\partial \bar\partial}{\partial^{+}}-x^{-}\partial-\lambda \frac{\partial}{\partial^{+}}) \ ,
\end{eqnarray} 
and their complex conjugates. We work on the surface $x^+=0\,$ to simplify calculations. The generators are of two kinds: kinematical that do not involve time derivatives 
\bea
P^+\;,\;\;P\;,\;\;\bar P\;,\;\; J \;,\;\;J^+\;,\;\;{\bar J}^+ \;\;{\mbox {and}}\;\; J^{+-}\ ,
\eea
and dynamical that do 
\bea
P^-\;,\;\;J^-\;,\;\;{\bar J}^-\ .
\eea
Dynamical generators pick up corrections when interactions are switched on. We introduce the Hamiltonian variation
\begin{eqnarray}
\label{hamvar}
\delta_{\mathcal H}\phi\equiv\partial^-\phi=\lbrace \phi,H\rbrace=\frac{\partial\bar\partial}{\partial^{+}}\,\phi\ ,
\end{eqnarray}
where the last equality only holds for the free theory. The appendix lists all non-vanishing commutators satisfied by the light-cone Poincar\'e generators. In the following, we adopt the approach of~\cite{BBB}: start with an ansatz for the operator $\delta_H\,\phi$, work through the list of Poincar\'e commutators to refine the ansatz and thus arrive at a structure for the Hamiltonian. For related discussions, see~\cite{Akshay}.

\vskip 0.3cm
\ndt Let us focus on the Hamiltonian describing $n$ fields. This would be at order $\kappa^{n-2}$ and have the schematic form $\kappa^{n-2}\;\phi^c\,\bar\phi^d$ ($c,d\in \{0,1,..,n\}$ and $c+d =n$). To arrive at such a Hamiltonian, we begin with the ansatz
\bea 
\delta_{\mathcal H}^{\kappa^{n-2}} \phi=\kappa^{n-2}\,{\partial^{+}}^{\mu_0}\biggl \{[{\bar\partial}^{{\alpha}_1} \,\partial^{{\beta}_1}\, {\partial^{+}}^{{\mu}_1}  \phi]\,...\,[{\bar\partial}^{{\alpha}_c} \,\partial^{{\beta}_c}\, {\partial^{+}}^{{\mu}_c}\phi]
\,\;[{\bar\partial}^{{\alpha}_{c+1}} \,\partial^{{\beta}_{c+1}}\, {\partial^{+}}^{{\mu}_{c+1}}\bar\phi]\,...\,
[{\bar\partial}^{{\alpha}_{n-1}} \,\partial^{{\beta}_{n-1}}\,{\partial^{+}}^{{\mu}_{n-1}} \bar\phi] \biggr\} \nn
\eea
where ${\alpha}_i$, $ {\beta}_i$, ${\mu}_i$ are integers to be fixed by the algebra. The next step is to work through some of the commutators in the appendix. The commutator $[\delta_{J},\delta_{H}]$ yields
\bea
\label{delj}
\sum_{i=1}^{n-1}({\alpha}_i - {\beta}_i) + (d-c)\lambda = 0\ .
\eea
The next commutator, $[\delta_{J^{+-}},\delta_{H}]$ produces the condition
\bea
\label{deljpm}
\sum_{i=0}^{n-1} {\mu}_i  = -1\ .
\eea
Using (\ref{deljpm}) and  dimensional analysis we find (noting that $[\kappa] = [L]^{\lambda -1}$) 
\bea
\label{dim}
\sum_{i=1}^{n-1} ({\alpha}_i + {\beta}_i) = 2 + (\lambda - 2)(c + d - 2)\ .
\eea
If we assume that there are no negative powers of the transverse derivatives in the Hamiltonian, using (\ref{delj}) and (\ref{dim})   we obtain 
\bea
\label{ineq1}
\sum_{i=1}^{n-1} {\alpha}_i = (\lambda - 1)c - d - \lambda + 3 \geq 0\ ,\\
\label{ineq2}
\sum_{i=1}^{n-1} {\beta}_i = (\lambda - 1)d - c - \lambda + 3 \geq 0\ .
\eea

\ndt Consider the implications of these inequalities in the simplest case: for spin-1 fields. $\lambda=1$ implies $c\,,d\,\leq \,2$ from which it follows that $n\leq 4$. Thus verifying the known result, that the Yang-Mills theory action terminates with four-point interaction vertices.
\vskip 0.3cm
\ndt Now we consider the case of gravity with $\lambda=2$. Expressions (\ref{ineq1}) and (\ref{ineq2}) imply that
\bea
 \label{gravineq1}
 \sum_{i=1}^{n-1} {\alpha}_i = c - d + 1 \geq 0\ ,\\
 \label{gravineq2}
 \sum_{i=1}^{n-1} {\beta}_i = d - c + 1 \geq 0\ .
 \eea
\ndt For a MHV vertex, $d = 2$. From (\ref{gravineq1}) and (\ref{gravineq2}), we see that $1 \leq c \leq 3$. As $c \,+\, d \,=\, n$, it follows that $3 \leq n \leq 5$ for a MHV vertex. Exactly as demonstrated in section 2, MHV vertices arise from the cubic, quartic and quintic interaction vertices. Also, as verified explicitly by our calculations, the first entirely non-MHV vertices appear in the 6-point result. Using $c+d=n$ in (\ref{gravineq1}) and (\ref{gravineq2}) yields

\bea
\label{gravityineq}
\frac{n - 1}{2}\, \leq \,c\,,d\, \leq\, \frac{n + 1}{2}\ .
\eea

\ndt This relation allows us to extract information about the general form of even- and odd-point vertices. Consider an even point vertex, where $n\, = \,2m$ for some positive integer $m$. We find
\bea
m - \frac{1}{2}\, \leq \,c\,,d\, \leq \,m + \frac{1}{2}\ ,
\eea
which implies $c = d = m$. Thus, the only allowed structure at order $2m$ is $h^m\,\,\bar{h}^m$. Now consider an odd-point vertex, where $n\, =\, 2m\, +\, 1$ for some positive integer $m$. We find
\bea
m\, \leq\, c\,,d \,\leq\, m +1\ .
\eea
If $c=m$, then $d = m+1$ and consequently, odd-point vertices will have the structure $h^{m}\,\,\Bar{h}^{m+1}$. The other case ($c=m+1$) produces the structure $h^{m+1}\,\,\Bar{h}^{m}$. These are the only two allowed possibilities at odd order. Further, due to the reality of the Lagrangian, both structures appear together (as complex conjugates of one another).

\ndt This clearly demonstrates the fact that MHV terms do not occur naturally beyond the quintic interaction vertex in light cone gravity. Thus, MHV vertices at any order beyond $\kappa^3$, occur only through the canonical field redefinitions described in (\ref{redef}).

\vskip 1cm

	\section*{Acknowledgments}
	The work of SA is partially supported by a MATRICS grant - MTR/2020/000073 - of SERB. NB and AR are supported by a CSIR NET fellowship and a DAE-DISHA fellowship respectively.

\newpage	
\appendix
\section{Light-cone Poincar\'e algebra}

\ndt We define

\be
J^+\ =\ \frac{J^{+1}+iJ^{+2}}{\sqrt{2}}\ ,\quad \bar J^{+}\ =\ \frac{J^{+1}-iJ^{+2}}{\sqrt 2}\ ,\quad J\ =\ J^{12}\ .
\ee
\vskip 0.2cm
\ndt
The {\it {non-vanishing}} commutators of the Poincar\'e algebra are

\bea
&[P^-, J^{+-}]\ =\ -i P^- \ , \quad  &[P^-, J^+] = -i P\ , \quad [P^-, \bar J^+]\ =\ -i \bar P \nn \\ \nn\\
&[P^+, J^{+-}]\ = \ iP^+\ ,\quad  &[P^+, J^-]\ =\ -iP\ , \quad [P^+, \bar J^-]\ =\ -i \bar P \nn \\ \nn \\
&[P, \bar J^-]\ =\ -i P^-\ ,  \quad &[P, \bar J^+] \ =\ -iP^+\ , \quad [P, J]\ =\ P \nn \\ \nn \\
&[\bar P, J^-]\ = \ -i P^-\ , \quad &[\bar P, J^+]\ =\ -iP^+\ , \quad [\bar P, J]\ =\ -\bar P \nn \\ \nn \\
&[J^-, J^{+-}]\ =\ -i J^- \ , \quad &[J^-, \bar J^+]\ =\ iJ^{+-} +  J \ , \quad [J^-, J]\ =\ J^- \nn \\ \nn \\
&[\bar J^-, J^{+-}]\ = \ -i \bar J^- \ , \quad &[\bar J^-, J^+]\ =\ iJ^{+-} - J \ , \quad [\bar J^-, J]\ =\ -\bar J^- \nn \\ \nn \\
&[J^{+-}, J^+]\ =\ -i J^+ \ ,  \quad &[J^{+-}, \bar J^+]\ = \ -i \bar J^+ \ , \nn \\ \nn \\
&[J^+, J]\ =\ J^+\ , \quad &[\bar J^+, J]\ =\ -\bar J^+\ .
\eea

\end{document}